\documentstyle[a4,12pt,epsf]{article}
\begin{document}

\title{Dynamics and thermodynamics of the pseudospin-electron model 
in the case of absence of the electron Hubbard correlation.
I. The analytical consideration.
}
\author{
I.V.Stasyuk, A.M.Shvaika, K.V.Tabunshchyk
}
\date{Institute for Condensed Matter Physics, Ukrainian National Academy of 
Sciences, 1 Svientsitskii St., 29011 Lviv, Ukraine}
\maketitle
\begin{abstract}
{Dynamics and thermodynamics of the model with local anharmonism
in the case of absence of the electron (Hubbard) correlation is investigated
in the present work. The correlation functions, mean values of pseudospin
and particle number as well as the thermodynamical potential are calculated.
The calculation is performed by diagrammatic method in the mean field
approximation. Single--particle Green functions are taken in the 
Hubbard--I approximation. }
\end{abstract}

\section{Introduction.}

The model considering the interaction of electrons with the local anharmonic 
mode of lattice vibrations is used in the last years in the theory of 
high--temperature superconducting crystals. Particularly, such property is
characteristic for the vibrations of the so--called apex oxygen ions
O$_{IV}$ along $c$--axis direction of the layered compounds of
YBa$_2$Cu$_3$O$_7$--type structure (see, [1-3]). An important role of the
apex oxygen and its anharmonic vibrations in the phase transition into
superconducting state has been already mentioned [4,5] and the possible
connection between the superconductivity and lattice instability of
ferroelectric type in high--$T_c$ superconducting compounds is under
discussion [6,7]. In the case of local double--well potential, the
vibrational degrees of freedom can be presented by pseudospin variables.
The Hamiltonian of the derived in this way pseudospin--electron model has
the following form [8]
\begin{equation}
\label{1.1}
H=\sum_iH_i+\sum_{ij\sigma}t_{ij}b^+_{i\sigma}b_{j\sigma}\hspace*{1ex},
\end{equation}
and includes besides the terms describing electron transfer ($\sim t_{ij}$),
the electron correlation ($U$--term), interaction with anharmonic mode
($g$--term), the energy of the tunnelling splitting ($\Omega$--term)
and energy of the anharmonic potential asymmetry ($h$--term) in the
single--site part
\begin{equation}
\label{1.2}
H_i=Un_{i\uparrow}n_{i\downarrow}+E_o(n_{i\uparrow}+n_{i\downarrow})+
g(n_{i\uparrow}+n_{i\downarrow})S^z_i-\Omega S^x_i
-hS^z_i\hspace*{1ex}.
\end{equation}
Here, $E_o$ gives the origin for energies of the electron states at lattice
site ($E_o=-\mu$).

In this paper, which consists of two parts,
our aim is to obtain the expressions for correlation
functions which determine the dielectric susceptibility, mean values of
pseudospin and particle number operators as well as the thermodynamical
potential in the case $\Omega =0$ and absence of the Hubbard correlation
$U=0$.

In the second part of the paper we shell perform numerical calculations for
the analytical expressions obtained in the first part. We shall investigate
values of pseudospin and particle number operators with the change of the
asymmetry parameter $h$ ($T=const$) or with the change of temperature $T$
($h=const$) for the cases of the fixed chemical potential value
(regime $\mu =const$) and constant mean value particle number. We
shell present also some result for the case $T=0$. On the basis of the
obtained results the applicability of the Hubbard--I aproximate will be 
discussed.

\section{Hamiltonian and initial relations.}

We shall write the Hamiltonian of the model and the operators which
correspond to physical quantities in the second quantized form using the
operators of the electron creation (annihilation) at the site with the
certain pseudospin orientation
\begin{equation}
\label{2.1}
\begin{array}{cc}
a_{\sigma i}=b_{\sigma i}(1/2+S^z_i)\hspace*{1ex},  &
a^+_{\sigma i}=b^+_{\sigma i}(1/2+S^z_i)\hspace*{1ex},\\
\tilde{a}_{\sigma i}=b_{\sigma i}(1/2-S^z_i)\hspace*{1ex}, &
\tilde{a}^+_{\sigma i}=b^+_{\sigma i}(1/2-S^z_i)\hspace*{1ex}.
\end{array}
\end{equation}
Then we obtain the following expression for the initial Hamiltonian
\begin{equation}
\label{2.2}
\begin{array}{ll}
H=\sum\limits_i\{\varepsilon (n_{i\uparrow}+n_{i\downarrow})+
\tilde{\varepsilon}(\tilde{n}_{i\uparrow}+\tilde{n}_{i\downarrow})-hS^z_i\}+
&\\
\hspace*{6ex}+\sum\limits_{ij\sigma}t_{ij}(a^+_{i\sigma}a_{j\sigma}+
a^+_{i\sigma}\tilde{a}_{j\sigma}+
\tilde{a}^+_{i\sigma}a_{j\sigma}+\tilde{a}^+_{i\sigma}
\tilde{a}_{j\sigma})\hspace*{1ex}=
&\\
\hspace{5em}
=H_o+H_{int},
\end{array}
\end{equation}
where
\begin{equation}
\label{2.3}
\varepsilon =E_o+g/2 \, , \,
\tilde{\varepsilon} =E_o-g/2 \, ,
\end{equation}
are energies of the single--site states;
$H_o$ is the single--site (diagonal) term,
$H_{int}$ is the hopping terms.

The introduced operators satisfy following commutation rules
\begin{equation}
\label{2.4}
\begin{array}{cc}
\{\tilde{a}^+_{i\sigma},\tilde{a}_{j\sigma'}\}=
\delta_{ij}\delta_{\sigma\sigma'}(1/2-S^z_i)\hspace*{1ex}, &
\{\tilde{a}^+_{i\sigma},a_{j\sigma'}\}=0\hspace*{1ex},\\
\{a^+_{i\sigma},a_{j\sigma'}\}=
\delta_{ij}\delta_{\sigma\sigma'}(1/2+S^z_i)\hspace*{1ex}, &
\{a^+_{i\sigma},\tilde{a}_{j\sigma'}\}=0\hspace*{1ex}.\\
\end{array}
\end{equation}
In order to calculate pseudospin mean values we shall use the standard
representation of the statistical operator in form
\begin{equation}
\label{2.5}
e^{-\beta H}=e^{-\beta H_o}\hat{\sigma}(\beta) \, ,
\end{equation}
\begin{equation}
\label{2.6}
\hat{\sigma}(\beta)=T_{\tau}\exp\left\{-\int\limits_0^\beta
H_{int}(\tau) d\tau \right\}\, .
\end{equation}
Which gives the following expressions for $\langle S^z_l\rangle$
\begin{equation}
\label{2.7}
\langle S^z_l \rangle=\frac{1}{\langle\hat{\sigma}(\beta)\rangle_o}
\langle S^z_l \hat{\sigma}(\beta)\rangle_o=
\langle S^z_l \hat{\sigma}(\beta)\rangle_o^c\, .
\end{equation}
Here, the operators are given in the interaction representation
\begin{equation}
\label{2.8}
A(\tau)=e^{\tau H_o}Ae^{-\tau H_o}\, ;
\end{equation}
the averaging $\langle\dots\rangle_o$ is performed over the statistical
distribution with Hamiltonian $H_o$, and symbol $\langle\dots\rangle_o^c$
denotes the keeping of the connected diagrams.

\section{Perturbation theory for pseudospin mean values and diagram
technique.}

The expansion of the exponent in $(\ref{2.6})$ in powers of $H_{int}$
$(\ref{2.2})$ leads, after the substitution in equation $(\ref{2.7})$, to
the expression that has the form of the sum of infinite series with the terms
presented by the averages of the $T$--products of the electron creation
(annihilation) operators at the site with
the certain pseudospin orientation in the
interaction representation. The evaluation of such averages can be performed
using Wick's theorem.

In our case this theorem has some differences from the standart formulation.
Namelly, result of each pairing of operators $(\ref{2.1})$ contains an
operator factors, i.e.
\begin{eqnarray}
\label{3.3}
&\stackrel{\line(0,-1){.3}\vector(-1,0){1.7}\line(-1,0){.8}\line(0,-1){.3}\hspace{2em}\null}
{a_i(\tau')a_o^+(\tau)}
=\hat{g}(\tau'-\tau)\delta_{io}P^+_i , \quad
\stackrel{\line(0,-1){.3}\vector(-1,0){1.7}\line(-1,0){.8}\line(0,-1){.3}\hspace{2em}\null}
{\tilde{a}_i(\tau')\tilde{a}_o^+(\tau)}
=\tilde{g}(\tau'-\tau)\delta_{io}P^-_i ,
\\
\nonumber
&\stackrel{\line(0,-1){.3}\vector(1,0){1.7}\line(1,0){.8}\line(0,-1){.3}\hspace{2em}\null}
{a_o^+(\tau)a_i(\tau')}
=-\hat{g}(\tau'-\tau)\delta_{io}P^+_i,
\quad
\stackrel{\line(0,-1){.3}\vector(1,0){1.7}\line(1,0){.8}\line(0,-1){.3}\hspace{2em}\null}
{\tilde{a}_o^+(\tau)\tilde{a}_i(\tau')}
=-\tilde{g}(\tau'-\tau)\delta_{io}P^-_i .
\end{eqnarray}
Finally, this gives the possibility to express result in term of the
products of nonperturbated Green functions
\begin{equation}
\label{3.4}
\hspace{-0.5em}
\hat{g}_{io}(\tau-\tau')=\frac{\langle T_\tau a_i(\tau)a_o^+(\tau')\rangle_o}
{\langle \{a_ia_o^+\}\rangle_o}=e^{\varepsilon(\tau'-\tau)}
\delta_{oi} \left\{
\begin{array}{rl}
\frac{1}{1+e^{-\beta\varepsilon}} & :\, \tau>\tau'\, ,\\
\frac{-1}{1+e^{\beta\varepsilon}} & : \,\tau'>\tau\, ,
\end{array}
\right.
\end{equation}
$$
\hspace*{-2em}
\tilde{g}_{io}(\tau-\tau')=\frac{\langle T_\tau \tilde{a}_i(\tau)
\tilde{a}_o^+(\tau')\rangle_o}{\langle \{\tilde{a}_i
\tilde{a}_o^+\}\rangle_o}=
e^{\tilde{\varepsilon}(\tau'-\tau)}
\delta_{oi} \left\{
\begin{array}{rl}
\frac{1}{1+e^{-\beta\tilde{\varepsilon}}} & : \,\tau>\tau'\, ,\\
\frac{-1}{1+e^{\beta\tilde{\varepsilon}}} & : \,\tau'>\tau\, ,
\end{array}
\right.
$$
$$
\hspace{-7.2em}
\tilde{g}_{io}(\tau-\tau')=\tilde{g}(\tau-\tau')\delta_{io}\, ,\hspace{1em}
\hat{g}_{io}(\tau-\tau')=\hat{g}(\tau-\tau')\delta_{io}\, ,
$$
and averages of the certain number of the projection operators
\begin{equation}
P^+_i=\frac{1}{2}+S^z_i\, ,\hspace*{3em} P^-_i=\frac{1}{2}-S^z_i.
\end{equation}
Let us demonstrate this procedure for the case of evaluation of
$\langle S^z_l \rangle$, for one of the terms which appear in the
fourth order of the perturbation theory
\begin{eqnarray}
\label{3.6}
\lefteqn{
\int\limits_0^\beta \!d\tau_1 \int\limits_0^\beta \!d\tau_2
\int\limits_0^\beta \!d\tau_3 \int\limits_0^\beta \!d\tau_4
\sum_{iji_1j_1}\sum_{i_2j_2i_3j_3}t_{ij}t_{i_1j_1}t_{i_2j_2}t_{i_3j_3}\times
}
\\ \nonumber
&&\times\langle T_\tau S^z_l a^+_i(\tau_1) a_j(\tau_1)
\tilde{a}^+_{i_1}(\tau_2)a_{j_1}(\tau_2) a^+_{i_2}(\tau_3)
\tilde{a}_{j_2}(\tau_3)a^+_{i_3}(\tau_4) a_{j_3}(\tau_4) \rangle_o\, .
\end{eqnarray}

The stepwise pairing of the certain operator with the other ones gives the
possibility to reduce expression $(\ref{3.6})$ to the sum of the averages of
the smaller number of operators
$$
\vspace*{1em}
\hspace*{-3em}
\langle T_\tau S^z_l a^+_i(\tau_1) a_j(\tau_1) \tilde{a}^+_{i_1}(\tau_2)
a_{j_1}(\tau_2) a^+_{i_2}(\tau_3) \tilde{a}_{j_2}(\tau_3)
a^+_{i_3}(\tau_4) a_{j_3}(\tau_4) \rangle_o=
$$
$$
\hspace*{-2em}
=\langle T_{\tau} S^z_l
\stackrel{\line(0,-1){.3}\vector(1,0){11.4}\line(1,0){8.6}\line(0,-1){.3}\hspace{2.3em}\null}
{a^+_i(\tau_1) a_j(\tau_1) \tilde{a}^+_{i_1}(\tau_2)
a_{j_1}(\tau_2) a^+_{i_2}(\tau_3) \tilde{a}_{j_2}(\tau_3)
a^+_{i_3}(\tau_4) a_{j_3}(\tau_4)} \rangle_o+
$$
$$
\hspace*{-1.5em}
+\langle T_\tau S^z_l
\stackrel{\line(0,-1){.3}\vector(1,0){4.9}\line(1,0){3.4}\line(0,-1){.3}\hspace{2.3em}\null}
{a^+_i(\tau_1) a_j(\tau_1) \tilde{a}^+_{i_1}(\tau_2)
a_{j_1}(\tau_2) }
a^+_{i_2}(\tau_3) \tilde{a}_{j_2}(\tau_3)
a^+_{i_3}(\tau_4) a_{j_3}(\tau_4) \rangle_o=
$$
\vspace*{1em}
$$
\hspace*{-1em}
=-\hat{g}_{ij_3}(\tau_1-\tau_4)\langle T_\tau S^z_l P^+_{j_3} a_j(\tau_1)
\tilde{a}^+_{i_1}(\tau_2) a_{j_1}(\tau_2) a^+_{i_2}(\tau_3)
\tilde{a}_{j_2}(\tau_3) a^+_{i_3}(\tau_4) \rangle_o-
$$
\begin{equation}
\label{3.7}
\end{equation}
$$
\hspace*{-0.2em}
-\hat{g}_{ij_1}(\tau_1-\tau_2)\langle T_\tau S^z_l P^+_{j_1} a_j(\tau_1)
\tilde{a}^+_{i_1}(\tau_2)  a^+_{i_2}(\tau_3)
\tilde{a}_{j_2}(\tau_3) a^+_{i_3}(\tau_4) a_{j_2}(\tau_3)\rangle_o\, .
$$
\\ [1ex]
The successive applications of the pairing procedure for $(\ref{3.7})$
leads, finally, to
$$
-\hat{g}_{ij_1}\!(\tau_1\!-\!\tau_2) \tilde{g}_{i_1j_2}\!(\tau_2\!-\!\tau_3)
\hat{g}_{i_3j}\!(\tau_4\!-\!\tau_1) \hat{g}_{i_2j_1}\!(\tau_3\!-\!
\tau_2)
\langle T_\tau S^z_l P^+_j P^+_{j_1} P^-_{j_2} P^+_{j_3}
\rangle_o-
$$
$$
-\hat{g}_{ij_3}\!(\tau_1\!-\!\tau_4) \tilde{g}_{i_1j_2}\!(\tau_2\!-\!\tau_3)
\hat{g}_{i_2j}\!(\tau_3\!-\!\tau_1) \hat{g}_{i_3j_1}\!(\tau_4\!-\!\tau_2)
\langle T_\tau S^z_l P^+_j P^+_{j_1} P^-_{j_2} P^+_{j_3}
\rangle_o+
$$
$$
+\hat{g}_{ij_3}\!(\tau_1\!-\!\tau_4) \tilde{g}_{i_1j_2}\!(\tau_2\!-\!\tau_3)
\hat{g}_{i_2j_1}\!(\tau_3\!-\!\tau_2) \hat{g}_{i_3j}\!(\tau_4\!-\!\tau_1)
\langle T_\tau S^z_l P^+_j P^+_{j_1} P^-_{j_2} P^+_{j_3}
\rangle_o\, .
$$
\begin{equation}
\label{3.8}
\end{equation}
We introduce the diagrammatic notations
$$
{\epsfysize 1.5cm\epsfbox{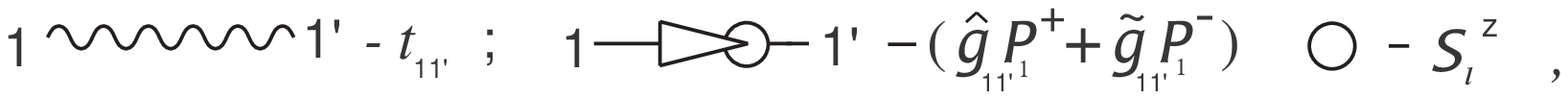}}
$$
and diagrams
$$
{\epsfysize 3.2cm\epsfbox{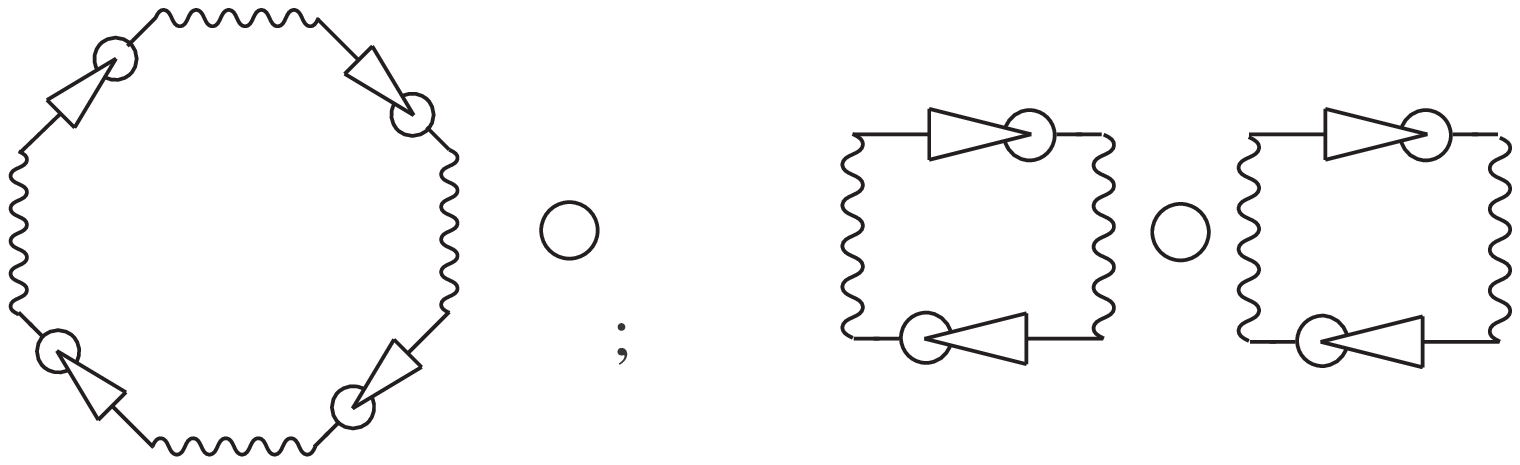}}
$$
correspond to the expression $(\ref{3.8})$.

The expansion of $(\ref{3.8})$ in semi--invariants leads to the
multiplication of diagrams (semi--invariants are represented by the ovals
surrounding corresponding vertices with diagonal operators
and contain the $\delta$--symbol on site indexes).
For example,
$$
\epsfxsize 0.9\textwidth\epsfbox{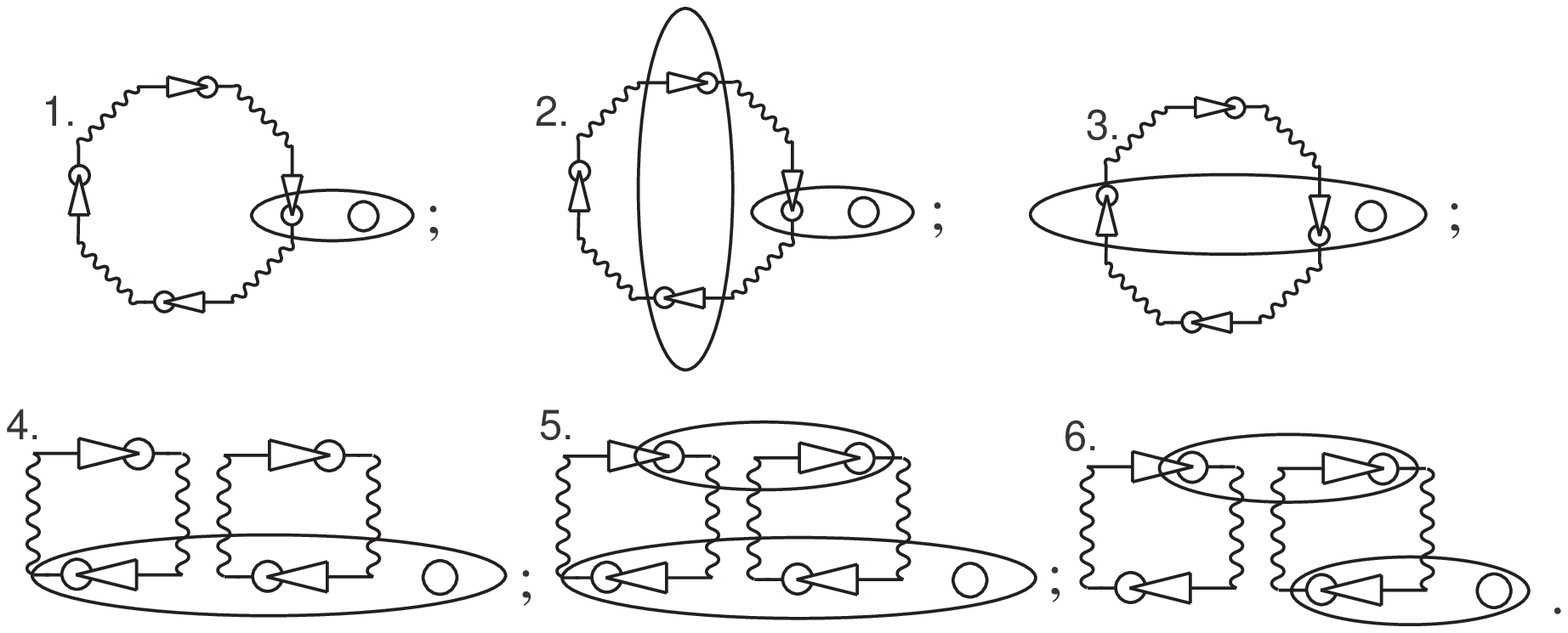}
$$
We shall neglect diagrams of types 2, 3, 5 i.e. the types including
semi--in\-va\-ri\-ants of the higher than first order in the loop (this
means that chain fragments form the single--electron Green functions in the
Hubbard--I approximation) and also the connection of
two loops by more than one
semi--invariant (this approximation means that selfconsistent field is taken
into account in the zero approximation).

Let us proceed to the momentum--frequency representation in the expressions
for the Green functions determined on finite interval $0{<}\tau{<}\beta$
when they can be expanded in Fourier series with discrete frequencies
$$
\hat{g}(\tau)=\frac{1}{\beta}\sum_n e^{i\omega_n \tau}g(\omega_n);
$$
\begin{equation}
\label{3.9}
\tilde{g}(\tau)=\frac{1}{\beta}\sum_n
e^{i\omega_n \tau}\tilde{g}(\omega_n);
\end{equation}
$$
\hat{g}(\omega_n)=-\frac{1}{i\omega_n-\varepsilon};\hspace*{3em}
\tilde{g}(\omega_n)=-\frac{1}{i\omega_n-\tilde{\varepsilon}};\hspace*{3em}
\omega_n=\frac{2n+1}{\beta}\pi\, .
$$
The characteristic feature of the already presented diagrams and diagrams
that correspond to other orders of the perturbation theory is the presence
of chain fragments. The simplest series of chain diagrams is
\begin{equation}
\label{3.10}
\raisebox{-1.1cm}[1.1cm][1.1cm]{\epsfysize 2.2cm\epsfbox{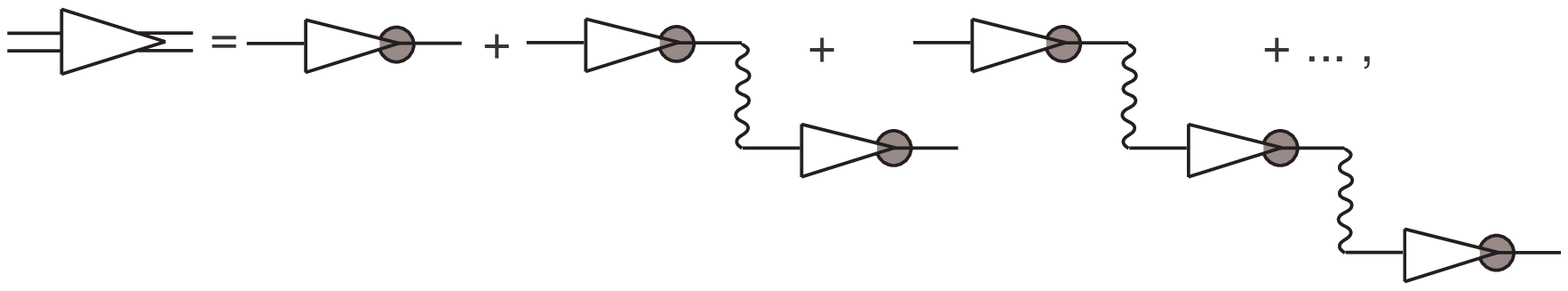}}
\end{equation}
where
\begin{equation}
\label{3.11}
\raisebox{-0.7cm}[0.8cm][0.7cm]{\epsfysize 1.5cm
\epsfbox{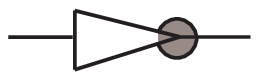}}\!\!\!\!\! =g(\omega_n)=\frac{\langle
P^+\rangle}{i\omega_n-\varepsilon}+
\frac{\langle P^-\rangle}{i\omega_n-\tilde{\varepsilon}}\, ,
\end{equation}
and corresponds to Hubbard--I approximation for single--electron
Green function. The expression
\begin{equation}
\label{3.12}
\raisebox{-0.7cm}[0.8cm][0.7cm]{\epsfysize 1.5cm\epsfbox{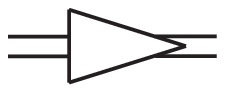}}\!\!\!\!\!
=G_{\vec{k}}(\omega_n)=\frac{1}{g^{-1}(\omega_n)-t_{\vec{k}}}\, ,
\end{equation}
in the momentum--frequency representation corresponds to the
sum of graphs
$(\ref{3.10})$. The poles of function $G_{\vec{k}}(\omega_n)$ determine the
spectrum of the single--electron excitations
\begin{equation}
\label{3.13}
\varepsilon_{I\! I,I}(t_{\vec{k}})=1/2(2E_o+t_{\vec{k}})\mp 1/2
\sqrt{g^2+4t_{\vec{k}}\langle S^z \rangle g +t^2_{\vec{k}}}\, ,
\end{equation}
Let us now return to the problem of summation of the diagram series for
average $\langle S^z_l\rangle$ taking into account the above mentioned
arguments. The diagram series has the form
\begin{equation}
\label{3.14}
\langle S^z_l\rangle=\!\!\!
\raisebox{-2.1cm}[1.4cm][2.1cm]{\epsfysize 3.5cm\epsfbox{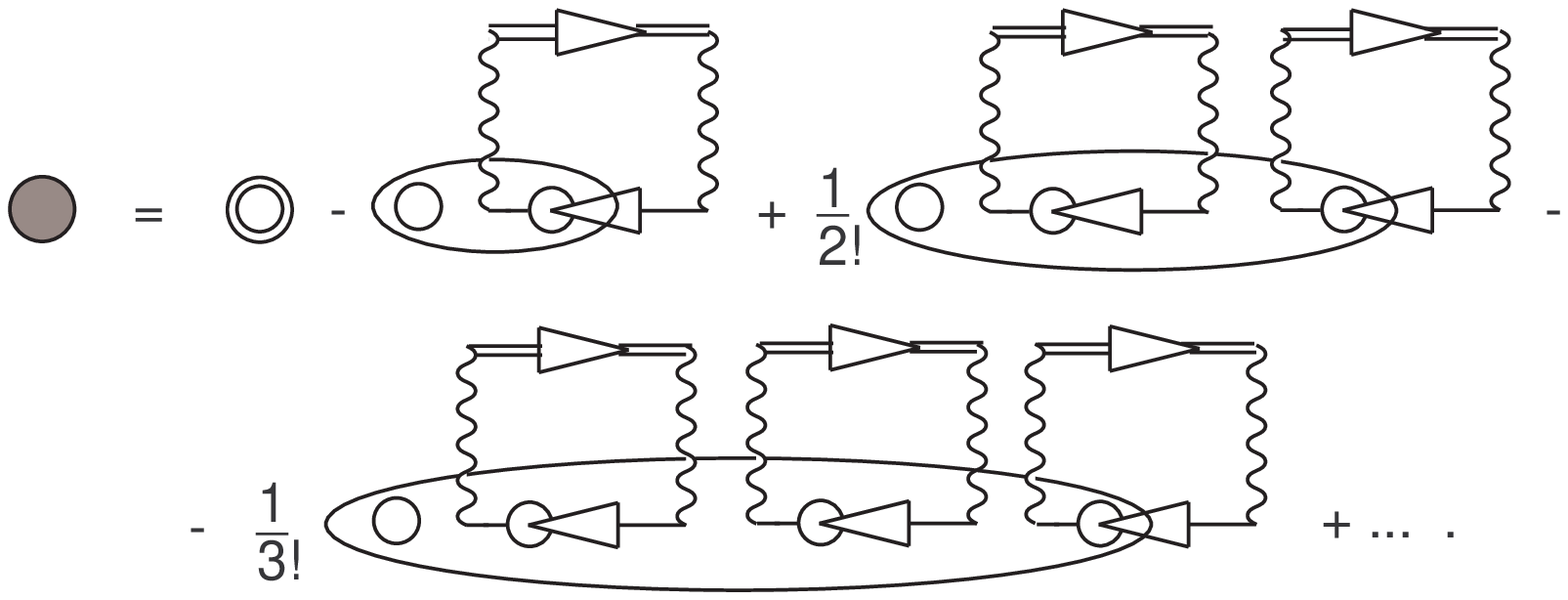}}
\end{equation}
The analytical expressions for loop has the next form
\begin{eqnarray}
\nonumber
\raisebox{-1cm}[1.2cm][1cm]{\epsfysize 2.2cm\epsfbox{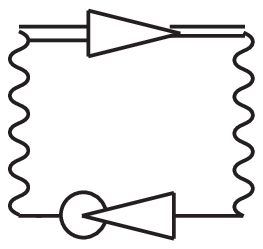}}
\! \! \! \! \! \!
&=&\frac{2}{N}\sum_{n,\vec{k}}\frac{t^2_{\vec{k}}}{g^{-1}(\omega_n)-
t_{\vec{k}}}\left(\frac{P^+_i}{i\omega_n-\varepsilon}+
\frac{P^-_i}{i\omega_n-\tilde{\varepsilon}}\right)=
\\
&=&\beta(\alpha_1P^+_i+\alpha_2P^-_i)\, ,
\end{eqnarray}
where we used the following notations
$$
\alpha_1=\frac{2}{N\beta}
\sum_{n,\vec{k}}\frac{t^2_{\vec{k}}}{(g^{-1}(\omega_n)-t_{\vec{k}})}
\frac{1}{(i\omega_n-\varepsilon)}\, ,
$$
$$
\alpha_2=\frac{2}{N\beta}
\sum_{n,\vec{k}}\frac{t^2_{\vec{k}}}{(g^{-1}(\omega_n)-t_{\vec{k}})}
\frac{1}{(i\omega_n-\tilde{\varepsilon})}\, .
$$
Using decomposition into simple fractions and summation over frequency we
obtained
$$
\alpha_1=\frac{2}{N}\sum_{\vec{k}}t_{\vec{k}}
(A_1n(\varepsilon_I(t_{\vec{k}}))+B_1n(\varepsilon_{I\! I}(t_{\vec{k}})))\, ,
$$
$$
\alpha_2=\frac{2}{N}\sum_{\vec{k}}t_{\vec{k}}
(A_2n(\varepsilon_I(t_{\vec{k}}))+B_2n(\varepsilon_{I\! I}(t_{\vec{k}})))\, ,
$$
where
$$
A_1=\frac{\varepsilon_I(t_{\vec{k}})-\tilde{\varepsilon}}
{\varepsilon_I(t_{\vec{k}})-\varepsilon_{I\! I}(t_{\vec{k}})}\, ,
\hspace*{4em}
B_1=\frac{\varepsilon_{I\! I}(t_{\vec{k}})-\tilde{\varepsilon}}
{\varepsilon_{I\! I}(t_{\vec{k}})-\varepsilon_I(t_{\vec{k}})}\, ,
$$
$$
A_2=\frac{\varepsilon_I(t_{\vec{k}})-\varepsilon}
{\varepsilon_I(t_{\vec{k}})-\varepsilon_{I\! I}(t_{\vec{k}})}\, ,
\hspace*{4em}
B_2=\frac{\varepsilon_{I\! I}(t_{\vec{k}})-\varepsilon}
{\varepsilon_{I\! I}(t_{\vec{k}})-\varepsilon_I(t_{\vec{k}})}\, .
$$
The equation for $\langle S^z_l\rangle$ can be presented in the form
\begin{eqnarray}
\nonumber
\lefteqn{
\langle S^z_l\rangle=\langle S^z_l\rangle_o-\langle S^z_l
\beta(\alpha_1P^+_l+\alpha_2P^-_l)\rangle_{oc}+}\\
\nonumber
&&+\frac{1}{2!}\langle S^z_l
\beta^2(\alpha_1P^+_l+\alpha_2P^-_l)^2\rangle_{oc}-\dots
=\langle S^z_le^{-\beta(\alpha_1P^+_l+\alpha_2P^-_l)}\rangle_{oc}\, .
\end{eqnarray}
Let us introduce
$$
H_{MF}=\sum\limits_iH_i^{MF}\, ,
$$
where
$$
H_i^{MF}=H_{i\,o}+\alpha_1P^+_i+\alpha_2P^-_i\, .
$$
Then the analytical equation for $\langle S^z_l\rangle$ can be expresed in
the form
\begin{eqnarray}
\label{3.15}
\lefteqn{
\langle S^z_l\rangle=\langle S^z_l\rangle_{MF}=
\frac{Sp(S^z_le^{-\beta H_{MF}})}{Sp(e^{-\beta H_{MF}})}=}
\\
\nonumber
&&=\frac{1}{2}\tanh\left\{\frac{\beta}{2}(h+\alpha_2-\alpha_1)+
\ln{\frac{1+e^{-\beta\varepsilon}}{1+e^{-\beta\tilde{\varepsilon}}}}
\right\}\, .
\end{eqnarray}
Difference $\alpha_2-\alpha_1$ corresponds to the internal effective
self--consistent field acting on pseudospin
\begin{equation}
\label{3.16}
\alpha_2-\alpha_1=\frac{2}{N}\sum_{\vec{k}}t_{\vec{k}}
\frac{\varepsilon-\tilde{\varepsilon}}
{\varepsilon_I(t_{\vec{k}})-\varepsilon_{I\! I}(t_{\vec{k}})}
\{n(\varepsilon_{I\! I}(t_{\vec{k}}))-n(\varepsilon_{I}(t_{\vec{k}})) \}\, .
\end{equation}

\section{Mean value of particle number.}

The diagram series for average $\langle n_i\rangle$ (using the perturbation
theory, Wick's theorem and expansion in semi--invariants) can be presented
in the form
\begin{equation}
\label{4.1}
\langle n_i\rangle
=\!\!\!\!\!\!\!\!
\raisebox{-2.5cm}[1.3cm][2.5cm]{\epsfysize 3.8cm\epsfbox{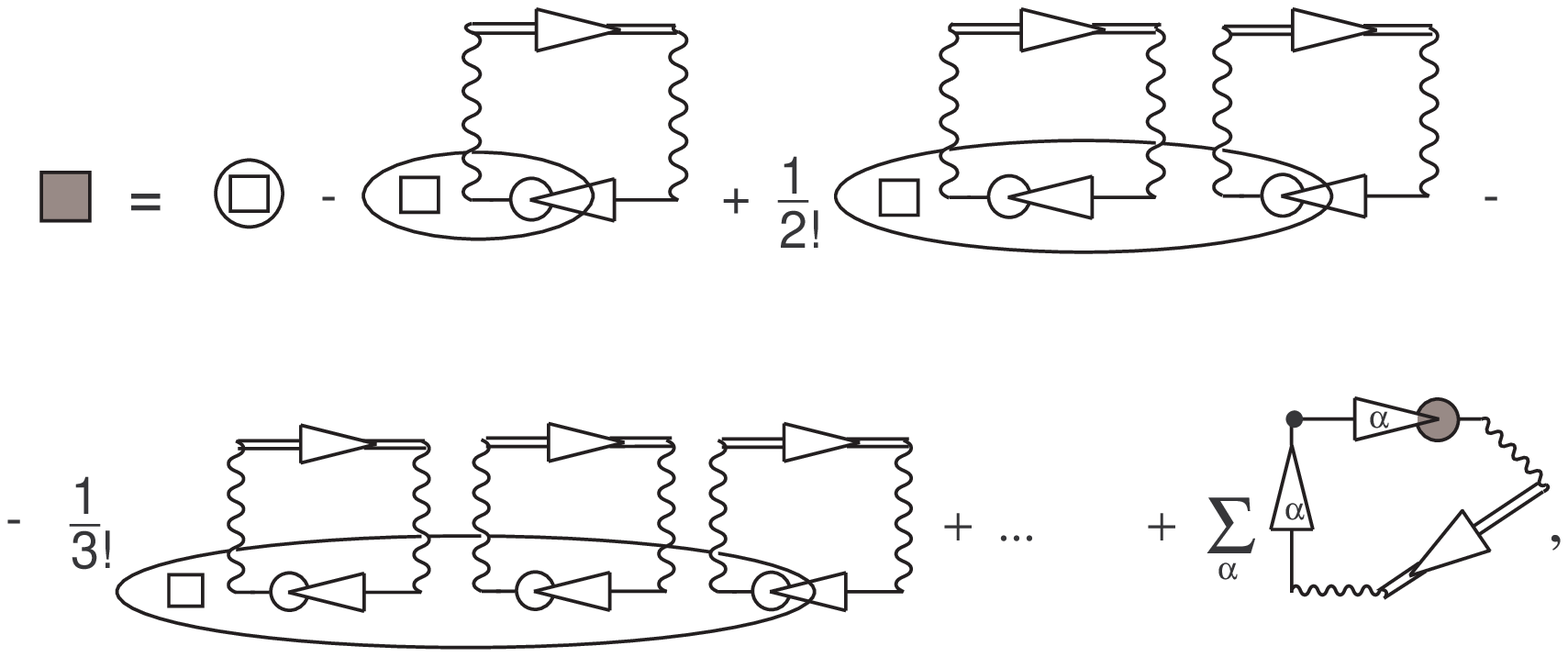}}
\end{equation}
where
$$
\raisebox{-0.7cm}[0.8cm][0.7cm]{\epsfysize 1.5cm\epsfbox{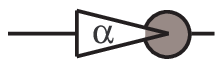}}\!\!\!\!
=\frac{\langle P^{\alpha}\rangle}{i\omega_n-\varepsilon^{\alpha}}\, ,
$$
$$
\raisebox{-0.7cm}[0.8cm][0.7cm]{\epsfysize 1.5cm\epsfbox{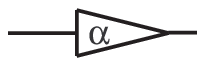}}\!\!\!\!
=\frac{1}{i\omega_n-\varepsilon^{\alpha}}\, , \,\,
\raisebox{-0.7cm}[0.8cm][0.7cm]{\epsfysize 1.5cm\epsfbox{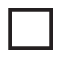}}\!\!\!\!
=\hat{n}_i\, , \,\,
$$
$$
P^{\alpha}=(P^+;P^-)\, ,
$$
$$
\varepsilon_{\alpha}=(\varepsilon;\tilde{\varepsilon})\, ,
$$
and the last term appears due to the pairing of the electron creation
(annihilation) operators with the operator of particle number.

Analytical expression for $(\ref{4.1})$ can be obtained starting from
formulas $(\ref{3.10})$, $(\ref{3.11})$
\begin{equation}
\label{4.2}
\langle n_i \rangle=\langle n_i \rangle_{MF}+\frac{2}{N\beta}
\sum_{n,\vec{k},\alpha}\frac{t^2_{\vec{k}}}{(g^{-1}(\omega_n)-t_{\vec{k}})}
\frac{\langle P^{\alpha}\rangle}{(i\omega_n-\varepsilon^{\alpha})^2}\, ,
\end{equation}
where
$$
\langle n_i \rangle_{MF}=\frac{Sp(n_ie^{-\beta H_{MF}})}
{Sp(e^{-\beta H_{MF}})}\, .
$$
After simple transformation we obtain next relation
\begin{equation}
\label{4.3}
\langle n \rangle_{MF}-n(\varepsilon)-n(\tilde{\varepsilon})=
2\langle S^z \rangle
(n(\varepsilon)-n(\tilde{\varepsilon}))\, ,
\end{equation}
or
$$
\langle n \rangle_{MF}=2\langle P^+\rangle
n(\varepsilon)+2\langle P^-\rangle
n(\tilde{\varepsilon})\, ,
$$
where $n(\varepsilon)=\frac{\displaystyle 1}
{\displaystyle 1+e^{\beta\varepsilon}}$
is Fermi distribution.

Using decomposition into simple fractions, summation over frequency and
relation $(\ref{4.3})$ we can present average $\langle n\rangle$ in the
form
\begin{equation}
\label{4.4}
\langle n \rangle=\frac{2}{N}
\sum_{\vec{k}}\{n(\varepsilon_I(t_{\vec{k}}))
+n(\varepsilon_{I\! I}(t_{\vec{k}})) \}-2\langle P^+\rangle
n(\tilde{\varepsilon})-2\langle P^-\rangle
n(\varepsilon)\, .
\end{equation}

\section{Thermodynamical potential.}

In order to calculate the thermodynamical potential let us introduce the
parameter $\lambda\in [0,1]$ in the initial Hamiltonian
\begin{equation}
\label{5.1}
H_{\lambda}=H_o+\lambda H_{int}\, ,
\end{equation}
such that $H\to H_o$ for
$\lambda=0$ and $H\to H_o+H_{int}$ for
$\lambda=1$.

Hence
$$
Z_{\lambda}=Sp(e^{-\beta H_{\lambda}})=
Sp(e^{-\beta H_o}\hat{\sigma}_{\lambda}(\beta))=
Z_o\langle\hat{\sigma}_{\lambda}(\beta)\rangle_o\, ,
$$
where
$$
\hat{\sigma}_{\lambda}(\beta)=T_{\tau}\exp\left\{\!-\lambda\int\limits_0^\beta
\! H_{int}(\tau) d\tau \right\}\, ,
$$
and
\begin{equation}
\label{5.2}
\Omega_{\lambda}=-\frac{1}{\beta}\ln Z_o
-\frac{1}{\beta}\ln\langle\hat{\sigma}_{\lambda}(\beta)\rangle_o\, ,
\end{equation}
$$
\Delta\Omega_{\lambda}=\Omega_{\lambda}-\Omega_o=
-\frac{1}{\beta}\ln\langle\hat{\sigma}_{\lambda}(\beta)\rangle_o\, .
$$
Here $\Omega_o$ is the thermodynamical potential calculated with the
single--site (diagonal) part of the initial Hamiltonian.

Therefore
\begin{equation}
\label{5.3}
\Delta\Omega=\int\limits_0^1 d\lambda
\! \left(\! \frac{d\Omega_{\lambda}}{d\lambda}\! \right)\, .
\end{equation}
For the value $d\Omega_{\lambda}/d\lambda$, we can write immediately the
diagram series in the next form
\begin{equation}
\label{5.4}
\beta \frac{d\Omega_{\lambda}}{d\lambda}=
\!\!\!\!\!
\raisebox{-1.2cm}[1.3cm][1.2cm]{\epsfysize 2.5cm\epsfbox{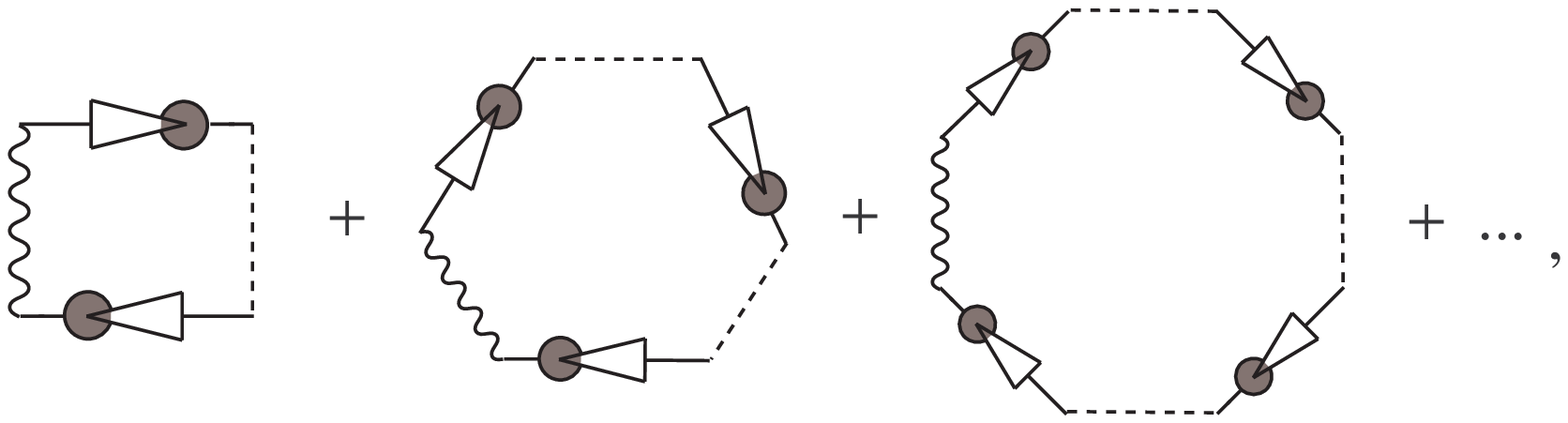}}
\end{equation}
where \hspace{3em}
\raisebox{-0.9cm}[1.1cm][0.9cm]{\epsfysize 2cm\epsfbox{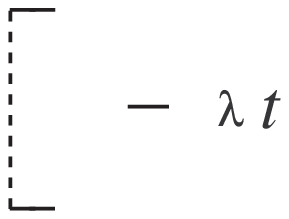}}\,
and also
$$
{\epsfysize 4cm\epsfbox{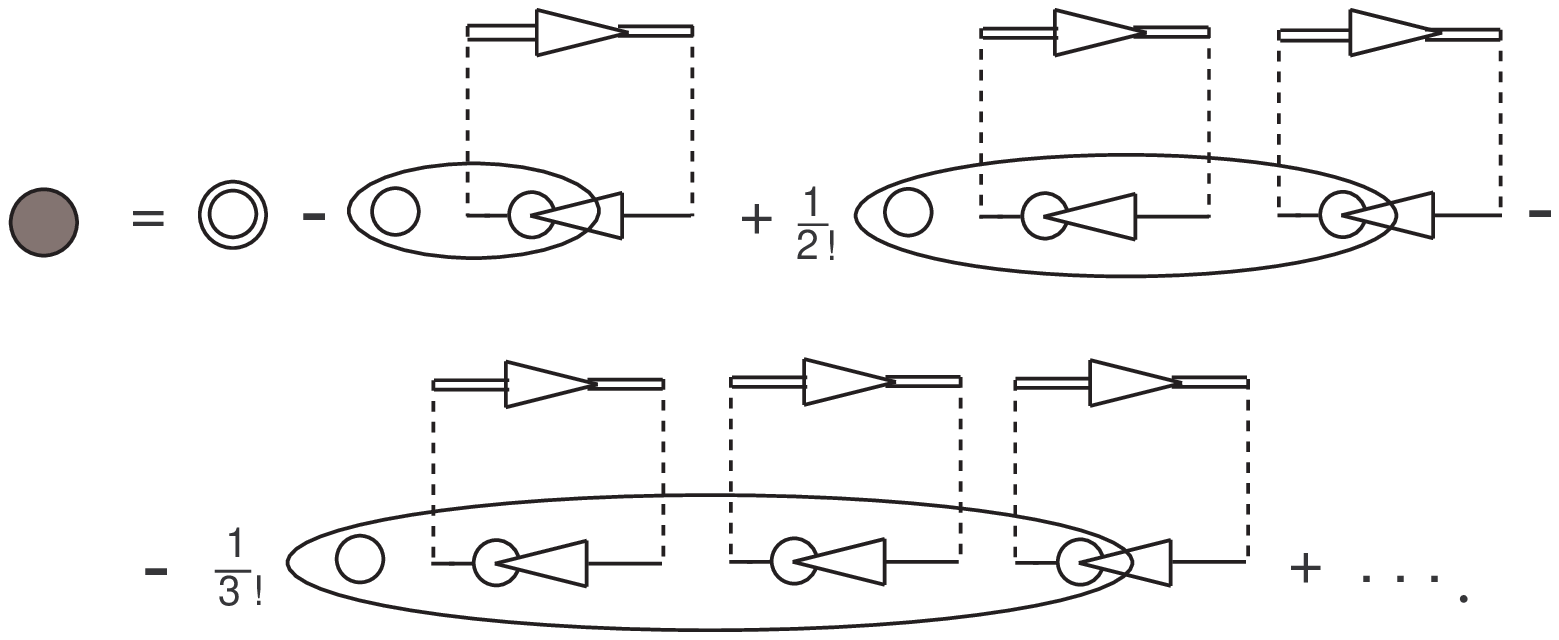}}
\hspace{-1.3in}\null
$$

The expression $(\ref{5.3})$ can be presented in the form (using the diagram
series $(\ref{5.4})$)
$$
\hspace*{-10em}
\Delta\Omega=\frac{2}{N\beta}\sum_{n,\vec{k}}
\int\limits_0^1
\lambda t^2_{\vec{k}}g^2_{\lambda}(\omega_n)
\frac{1}{1-\lambda t_{\vec{k}}g_{\lambda}(\omega_n)}
d\lambda=
$$
\begin{equation}
\label{5.5}
=-\frac{2}{N\beta}\sum_{n,\vec{k}}\ln(1-t_{\vec{k}}g(\omega_n))-
\frac{2}{N\beta}\sum_{n,\vec{k}}\int\limits_0^1
\frac{\lambda t_{\vec{k}}\frac{dg_{\lambda}(\omega_n)}{d\lambda}}
{1-\lambda t_{\vec{k}}g_{\lambda}(\omega_n)} d\lambda\, .
\end{equation}
The first term in expression $(\ref{5.5})$ may be written in the diagram
form as
\begin{equation}
\label{5.6}
\raisebox{-1.5cm}[1.5cm][1.5cm]{\epsfysize 3cm\epsfbox{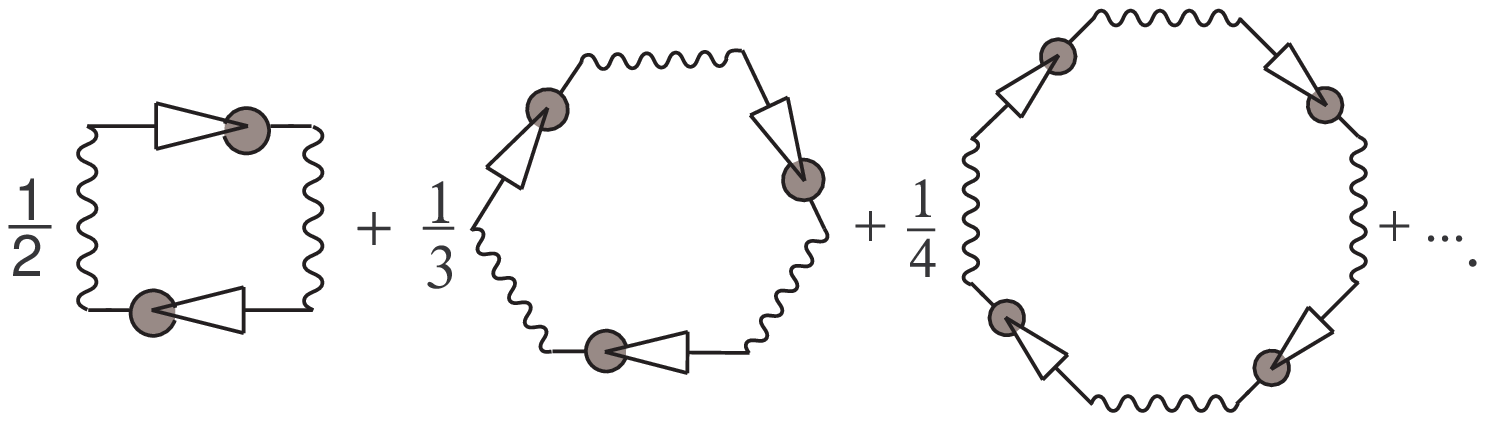}}
\hspace{-1.5in}\null
\end{equation}
The series $(\ref{5.6})$ describes an electron gas which
energy spectrum is
defined by the total pseudospins field. This series is in conformity with
the so--called one loop--approximation.

The second term in expression $(\ref{5.5})$ can be integrated to the
following diagram series
\begin{equation}
\label{5.7}
\null\hspace{-.7in}
\raisebox{-1.5cm}[1.5cm][1.5cm]{\epsfysize 3cm\epsfbox{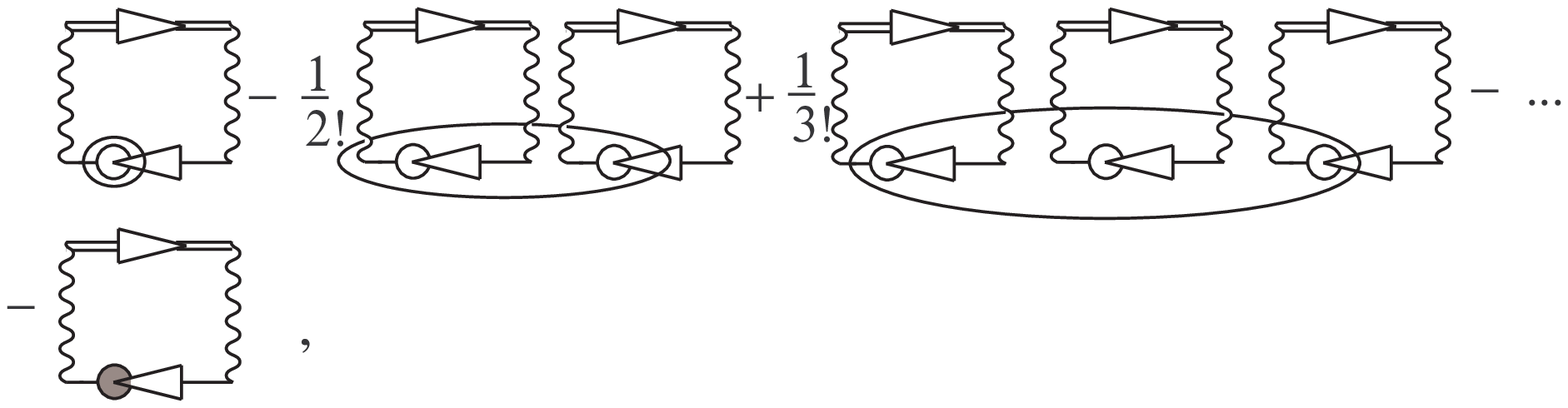}}
\hspace{-.8in}\null
\end{equation}
and appears due to the presence of pseudospin subsystem.

Finally, the diagram series for $\beta\Delta\Omega $ may be written as
the sum of expressions $(\ref{5.6})$ and $(\ref{5.7})$,
the corresponding analytical expression is the following
\begin{equation}
\label{5.9}
\Delta\Omega=-\frac{2}{N\beta}\sum_{\vec{k}}
\ln\frac{(\cosh\frac{\beta}{2}\varepsilon_I(t_{\vec{k}}))
(\cosh\frac{\beta}{2}\varepsilon_{I\! I}(t_{\vec{k}}))}
{(\cosh\frac{\beta}{2}\varepsilon)
(\cosh\frac{\beta}{2}\tilde{\varepsilon})}-
\end{equation}
$$
\hspace{-1em}
-\frac{1}{\beta}\ln \cosh\left\{
\frac{\beta}{2}(h+\alpha_2-\alpha_1)+\ln\frac
{1+e^{-\beta\varepsilon}}{1+e^{-\beta\tilde{\varepsilon}}}
\right\}+
$$
$$
+\frac{1}{\beta}\ln \cosh\left\{
\frac{\beta}{2} h+\ln\frac
{1+e^{-\beta\varepsilon}}{1+e^{-\beta\tilde{\varepsilon}}}
\right\}+\langle S^z\rangle(\alpha_2-\alpha_1)\, .
$$
Here, the decomposition in simple fractions and summation over
frequency were done.

Then, since the thermodynamical potential is the function of argument
$\langle S^z\rangle$, let us check the consistency of approximations made
for $\langle S^z\rangle$, $\langle n\rangle$ and thermodynamical potential
$\Omega$.
For this purpose one should obtain average
$\langle S^z\rangle$ and average $\langle n\rangle$ from the expression
for grand thermodynamical potential
$$
\frac{d\Omega}{d(-\mu)}=\frac{2}{N}
\sum_{\vec{k}}\{n(\varepsilon_I(t_{\vec{k}}))
+n(\varepsilon_{I\! I}(t_{\vec{k}})) \}-2\langle P^+\rangle
n(\tilde{\varepsilon})-2\langle P^-\rangle
n(\varepsilon)\, ,
$$
$$
\frac{d\Omega}{d(-h)}=
\frac{1}{2}\tanh\left\{\frac{\beta}{2}(h+\alpha_2-\alpha_1)+
\ln{\frac{1+e^{-\beta\varepsilon}}{1+e^{-\beta\tilde{\varepsilon}}}}
\right\}\, .
$$
We thus obtain
$$
\frac{d\Omega}{d(-\mu)}=\langle n \rangle\, ,\hspace{4em}
\frac{d\Omega}{d(-h)}=\langle S^z \rangle\, .
$$
Therefore, the calculation of the mean values of the pseudospin and particle
number operators as well as the thermodynamical potential is performed in
the same approximation which corresponds to the mean field one.

\section{Pseudospin, electron, and mixed correlators.}

In this section our aim is to calculate correlators
$$
K^{ss}_{lm}(\tau-\tau')=
\langle T\tilde{S}^z_l(\tau)\tilde{S}^z_m(\tau')
\rangle_c\, ,
$$
$$
K^{sn}_{lm}(\tau-\tau')=
\langle T\tilde{S}^z_l(\tau)\tilde{n}_m(\tau')
\rangle_c\, ,
$$
$$
K^{nn}_{lm}(\tau-\tau')=
\langle T\tilde{n}_l(\tau)\tilde{n}_m(\tau')
\rangle_c
$$
constructed of the operators given in the Heisenberg representation with
imaginary time argument.

Let us present the diagram series for correlation function (in the
momentum--frequency representation) within the generalized random phase
approximation. In our case (absence of the Hubbard correlation)
this approximation is reduced, because the so--called ladder diagrams
(see.[10]) with antiparallel lines disappear. This reduce allow to take into
account mean values of pseudospin found self--consistently within the mean
field approximation.

We would like to remind that we have neglected diagrams which include
semi--invariants of the higher than first order in the loop and
also connection of two loops by more than one semi--invariant.
\begin{equation}
\label{6.1}
\raisebox{-1.2cm}[1.3cm][1.2cm]{\epsfysize 2.5cm\epsfbox{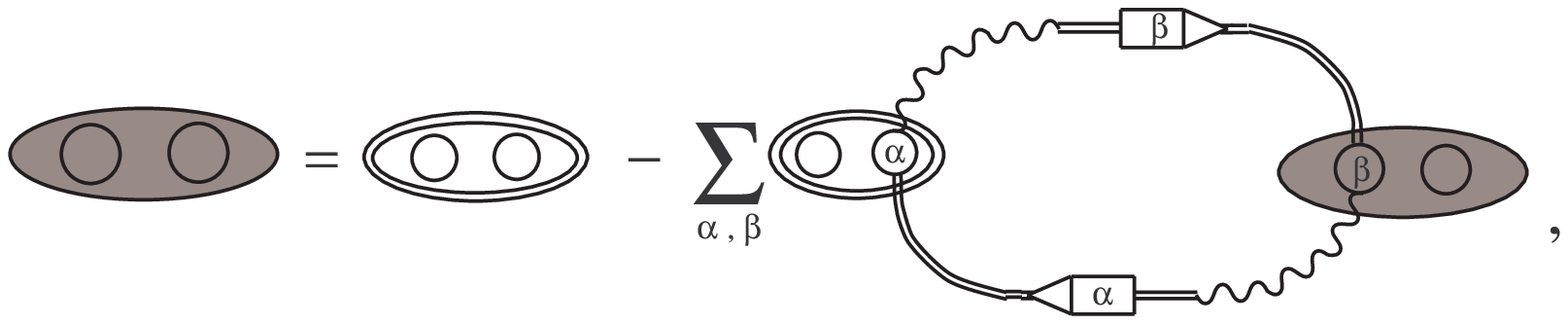}}
\end{equation}
where, we define
$$
\hspace{-2.8em}\,
\raisebox{-1.4cm}[0.7cm][1.4cm]{\epsfysize 2.1cm
\epsfbox{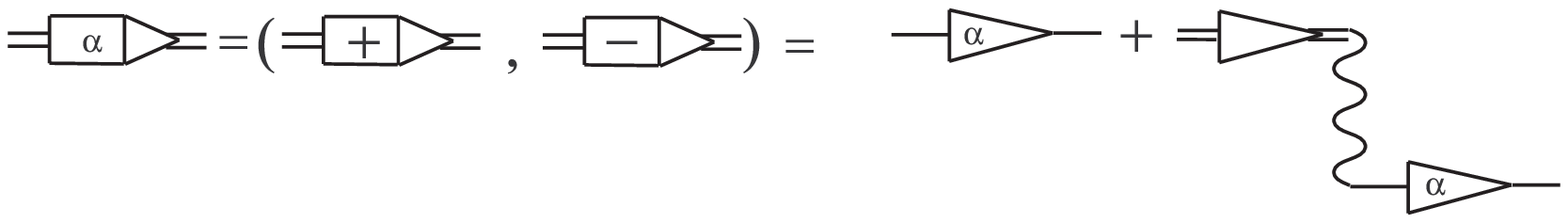}}\hspace{-4em}\!\!
=\! \Gamma^{\alpha}(\vec{k},\omega_n);
$$
$$
P^{\alpha}=(P^+,P^-);\,\,  \alpha=(0,1);\,\, \varepsilon^{\alpha}
=(\varepsilon,\tilde{\varepsilon});\,
\raisebox{-0.6cm}[0.9cm][0.6cm]{\epsfysize 1.5cm\epsfbox{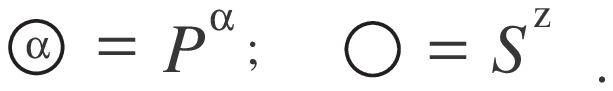}}
\hspace{-5mm}\null
$$
Here, the first term in equation $(\ref{6.1})$ takes into account the direct
influence on pseudospins of the internal effective self--consistent field
and is given by
\begin{equation}
\label{6.2}
\raisebox{-1.7cm}[1.8cm][1.7cm]{\epsfysize 3.5cm\epsfbox{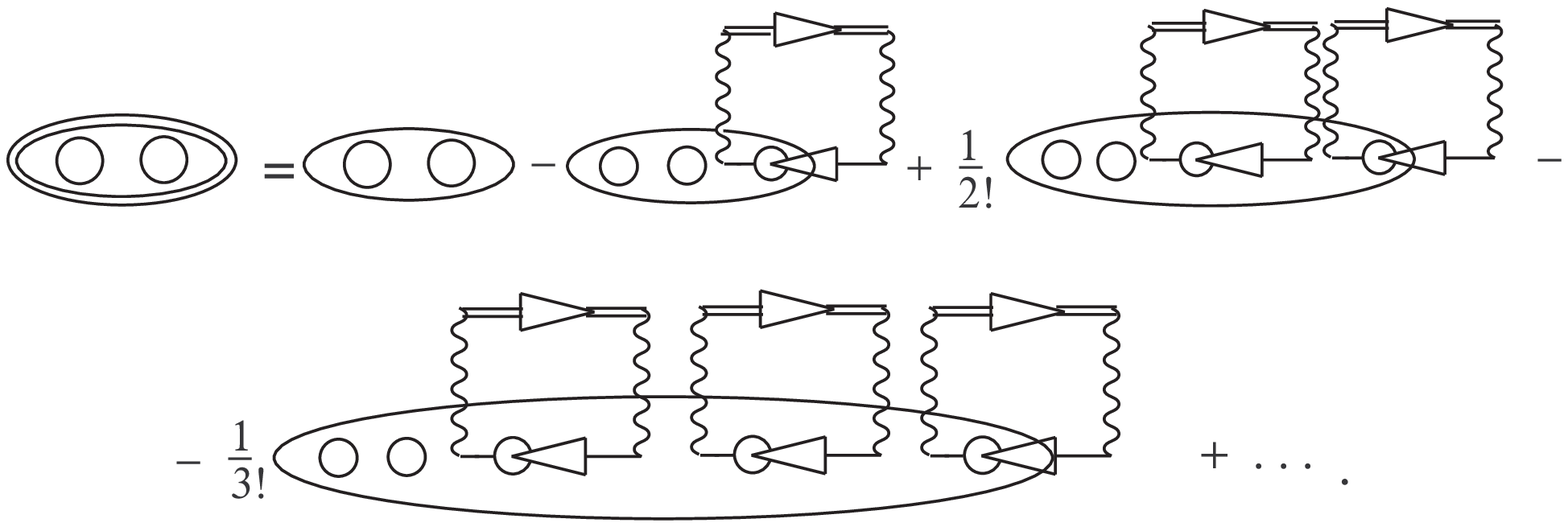}}
\end{equation}
Series $(\ref{6.2})$ means second--order semi--invariant which
renormalized due to the 'single--tale' parts, and thus is calculate
by $H_{MF}$.

The second term in equation $(\ref{6.1})$ describes the interaction between
pseudospins which is mediated by electrons (energy of electron spectrum is
defined by the total pseudospin field).

We introduce the shortened notations
\begin{equation}
\mbox{\fbox{$\Pi$}}^{\alpha,\beta}_{\vec{q}}=\!\!\!\!\!
\raisebox{-1.5cm}[1.5cm][1.5cm]{\epsfysize 3cm\epsfbox{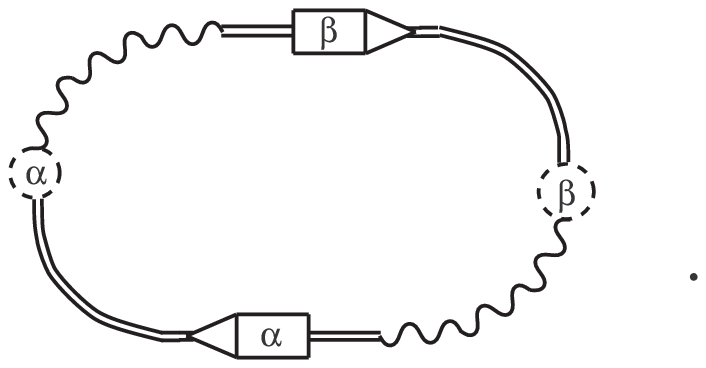}}
\label{6.3}
\end{equation}
Solution of equation $(\ref{6.1})$ can be written in the analytical form
\begin{equation}
\label{6.4}
\langle S^z S^z\rangle_{\vec{q}}=\frac{1/4-\langle S^z\rangle^2}
{1+\sum\limits_{\alpha,\beta}(-1)^{\alpha+\beta}
\mbox{\fbox{$\Pi$}}^{\alpha,\beta}_{\vec{q}}(\frac{1}{4}-\langle 
S^z\rangle^2)}\, , \end{equation}
where
\begin{equation}
\label{6.5}
\mbox{\fbox{$\Pi$}}^{\alpha,\beta}_{\vec{q}}=\frac{2}{N}\sum_{\vec{k}}t_{\vec{k}}
t_{\vec{k}+\vec{q}}\Gamma^{\alpha}(\vec{k},\omega_n)
\Gamma^{\beta}(\vec{k}+\vec{q},\omega_n)\, ,
\end{equation}
\begin{equation}
\label{6.6}
\Gamma^{\alpha}(\vec{k},\omega_n)=\frac{1}{(i\omega_n-\varepsilon^{\alpha})}
\frac{1}{(1-t_{\vec{k}}g(\omega_n))}\, .
\end{equation}
Decomposition of the function $\Gamma^{\alpha}(\vec{k},\omega_n)$ into simple
fractions and subsequent evaluation of the sum over frequency leads to the
next expression
$$
\sum_{\alpha,\beta}(-1)^{\alpha+\beta}
\mbox{\fbox{$\Pi$}}^{\alpha,\beta}_{\vec{q}}=\frac{2\beta}{N}
\sum_{\vec{k}} \frac
{t_{\vec{k}}t_{\vec{k}+\vec{q}}
(\varepsilon-\tilde{\varepsilon})^2}
{[\varepsilon_I(t_{\vec{k}})-\varepsilon_{I\! I}(t_{\vec{k}})]
[\varepsilon_I(t_{\vec{k}+\vec{q}})-\varepsilon_{I\! I}(t_{\vec{k}+\vec{q}})]}\times
$$
\begin{eqnarray}
\label{6.7}
\times\left\{\frac{n[\varepsilon_I(t_{\vec{k}})]-
n[\varepsilon_I(t_{\vec{k}+\vec{q}})]}
{\varepsilon_I(t_{\vec{k}})-\varepsilon_I(t_{\vec{k}+\vec{q}})}+
\frac{n[\varepsilon_{I\! I}(t_{\vec{k}})]-
n[\varepsilon_{I\! I}(t_{\vec{k}+\vec{q}})]}
{\varepsilon_{I\! I}(t_{\vec{k}})-\varepsilon_{I\! I}(t_{\vec{k}+\vec{q}})}-
\right.& & \\
\nonumber
\\ [1em]
\left.
-\frac{n[\varepsilon_I(t_{\vec{k}})]-n[\varepsilon_{I\! I}(t_{\vec{k}+\vec{q}})]}
{\varepsilon_I(t_{\vec{k}})-\varepsilon_{I\! I}(t_{\vec{k}+\vec{q}})}-
\frac{n[\varepsilon_{I\! I}(t_{\vec{k}})]-
n[\varepsilon_{I}(t_{\vec{k}+\vec{q}})]}
{\varepsilon_{I\! I}(t_{\vec{k}})-\varepsilon_{I}(t_{\vec{k}+\vec{q}})}
\right\}\, .& & 
\nonumber
\end{eqnarray}
After the substitution $(\ref{6.7})$ in equation $(\ref{6.4})$ we obtain,
finally, expression for $\langle S^zS^z\rangle$.

This formula for the uniform case $(\vec{q}=0)$ can be rewritten
as
\begin{equation}
\label{6.8}
\langle S^zS^z\rangle_{\vec{q}=0}=(1/4-\langle S^z\rangle^2)
\times
\end{equation}
$$
\times
\left\{1-\left(
\frac{4\beta}{N}\sum_{\vec{k}}t_{\vec{k}}^2
\frac{(\varepsilon-\tilde{\varepsilon})^2}
{[\varepsilon_I(t_{\vec{k}})-\varepsilon_{I\! I}(t_{\vec{k}})]^3}
\{n[\varepsilon_I(t_{\vec{k}})]-n[\varepsilon_{I\! I}(t_{\vec{k}})]\}+
\right.
\right.
$$
$$
\left.
\left.
+\frac{\beta^2}{2N} \sum_{\vec{k}}
\frac{t_{\vec{k}}^{2} (\varepsilon -\tilde{\varepsilon})^{2}}
{[\varepsilon_I(t_{\vec{k}}) -\varepsilon_{I\! I}(t_{\vec{k}})]^{2}}  
\left\{\frac{1}{\cosh^{2}\frac{\beta\varepsilon_I(t_{\vec{k}})}{2}}+ 
\frac{1}{\cosh^{2}\frac{\beta\varepsilon_{I\! I}(t_{\vec{k}})}{2}}  
\right\}\right) (\frac{1}{4} -\langle S^z\rangle^{2}) 
\right\}^{-1}.
$$ 
Expression $(\ref{6.8})$ can be obtained from the derivative $d\langle 
S^z\rangle/ d(\beta h)$.
This means that mean values of pseudospin and pseudospin correlators are
derived in the same approximation.

For mixed correlator the diagram series has the form
\begin{equation}
\label{6.9}
\langle S^zn\rangle\, =
\!\!\!\
\raisebox{-.8cm}[1.3cm][0.8cm]{\epsfysize 2.1cm\epsfbox{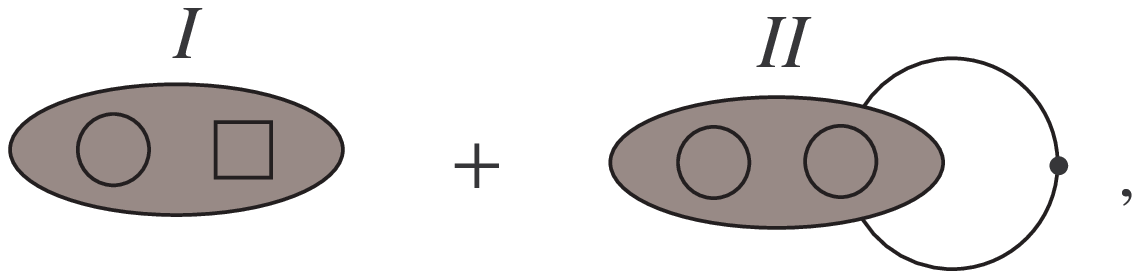}}
\end{equation}
where
\begin{equation}
\label{6.10}
\! I=\!
\!\!\!\!\!\!
\raisebox{-1.4cm}[1.4cm][1.4cm]{\epsfysize 2.8cm\epsfbox{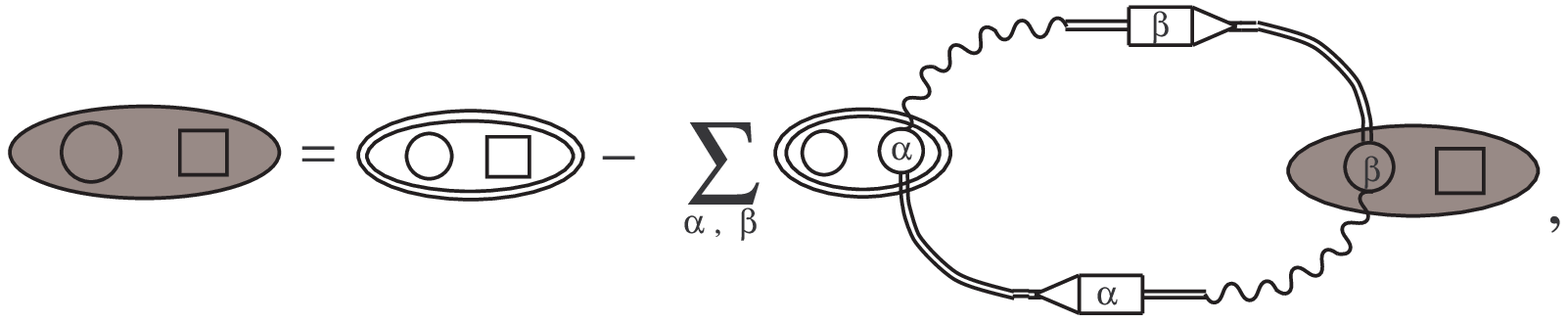}}
\hspace{-4mm}\null
\end{equation}
and
\begin{equation}
\label{6.11}
I\! I=
\!\!\!\!
\raisebox{-1.3cm}[1.8cm][1.3cm]{\epsfysize 3.1cm\epsfbox{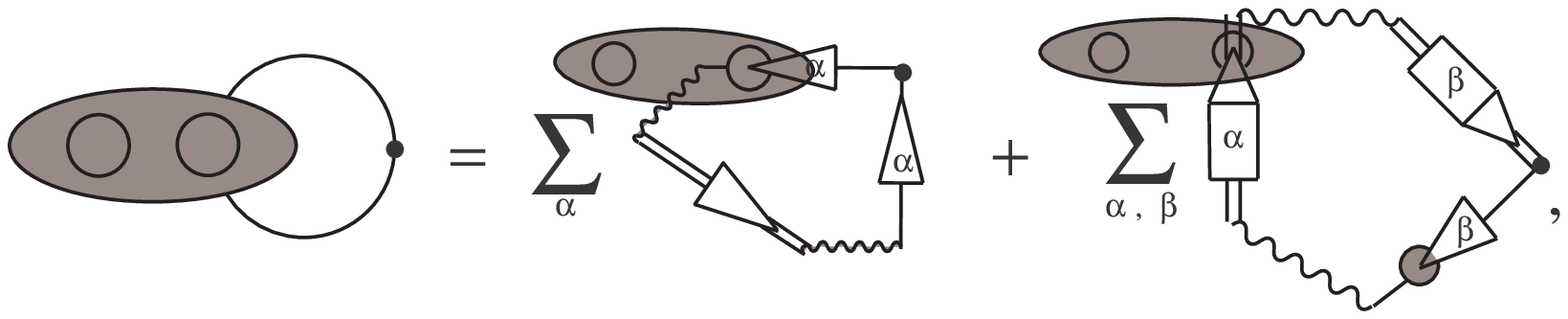}}
\hspace{-2mm}\null
\end{equation}
$$
\hspace{-2.6em}\,
\raisebox{-1.4cm}[0.7cm][1.4cm]{\epsfysize 2.1cm 
\epsfbox{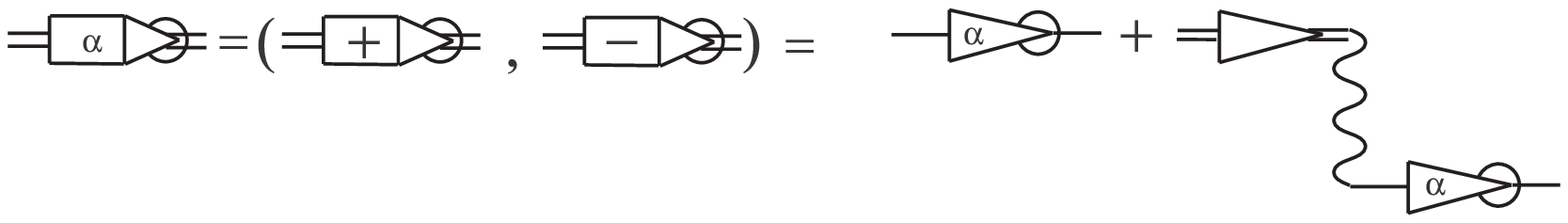}}\hspace{-4em}\! =\! 
P^{\alpha}\Gamma^{\alpha}(\vec{k},\omega_n)\, . $$
Solution of equation $(\ref{6.10})$ can be written in the analytical form
\begin{equation}
\label{6.12}
I=
2(n(\varepsilon)-n(\tilde{\varepsilon}))\langle S^z S^z\rangle_{\vec{q}}\, .
\end{equation}
Here we start from formula $(\ref{6.4})$ and from the next relation
\begin{equation}
\label{6.13}
\frac{\langle S^zn\rangle_{MF}-\langle S^z\rangle\langle n\rangle}
{\frac{1}{4}-\langle S^z \rangle^2}=
2(n(\varepsilon)-n(\tilde{\varepsilon}))\, .
\end{equation}

The second term in the diagram series $(\ref{6.9})$ we can present as
$$
I\! I=
\frac{2}{N}\langle S^z S^z\rangle_{\vec{q}}\sum_{\vec{k}}\frac
{t_{\vec{k}}(\varepsilon-\tilde{\varepsilon})}
{\varepsilon_I(t_{\vec{k}})-\varepsilon_{I\! I}(t_{\vec{k}})}\times
$$
\begin{eqnarray}
\label{6.14}
\times\left[\frac{n[\varepsilon_I(t_{\vec{k}})]-
n[\varepsilon_I(t_{\vec{k}+\vec{q}})]}
{\varepsilon_I(t_{\vec{k}})-\varepsilon_I(t_{\vec{k}+\vec{q}})}+
\frac{n[\varepsilon_{I}(t_{\vec{k}})]-
n[\varepsilon_{I\! I}(t_{\vec{k}+\vec{q}})]}
{\varepsilon_{I}(t_{\vec{k}})-\varepsilon_{I\! I}(t_{\vec{k}+\vec{q}})}
- \right. & &\\
\left.
-\frac{n[\varepsilon_{I\! I}(t_{\vec{k}})]-
n[\varepsilon_{I}(t_{\vec{k}+\vec{q}})]}
{\varepsilon_{I\! I}(t_{\vec{k}})-\varepsilon_{I}(t_{\vec{k}+\vec{q}})}-
\frac{n[\varepsilon_{I\! I}(t_{\vec{k}})]-
n[\varepsilon_{I\! I}(t_{\vec{k}+\vec{q}})]}
{\varepsilon_{I\! I}(t_{\vec{k}})-\varepsilon_{I\! I}(t_{\vec{k}+\vec{q}})}
\right]\, . & &
\nonumber
\end{eqnarray}
Let us introduce the shortened notations for the expression $(\ref{6.14})$
\begin{equation}
\label{6.15}
I\! I=\langle S^zS^z\rangle_{\vec{q}}\times [\oplus]_{\vec{q}}\, .
\end{equation}
In this way we obtain
\begin{equation}
\label{6.16}
\langle S^z n\rangle_{\vec{q}}=
2(n(\varepsilon)-n(\tilde{\varepsilon}))\langle S^z S^z\rangle_{\vec{q}}+
\langle S^z S^z\rangle_{\vec{q}}\times [\oplus]_{\vec{q}}\, .
\end{equation}
From our diagram series we can see: the correlators containing pseudospin
variable $S^z$ are different from zero only in the static case. This is due
to the fact that the operator $S^z$ commutes with Hamiltonian being the
integral of motion.

For electron correlator our diagram series has the form
\begin{equation}
\label{6.17}
\langle nn\rangle_{\vec{q},\omega}=
\!\!\!\!
\raisebox{-2.4cm}[1.1cm][2cm]{\epsfysize 3.5cm\epsfbox{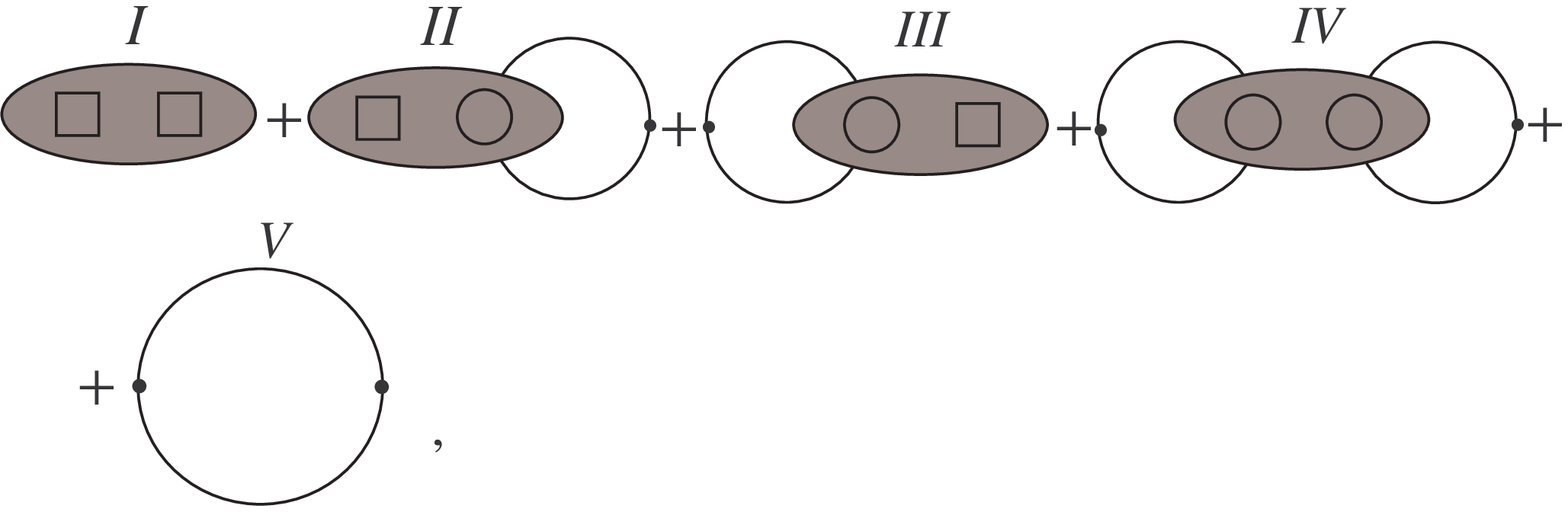}}
\hspace{-1mm}\null
\end{equation}
and only last term is not equal to zero for non--zero frequencies.
Let us consider series $(\ref{6.17})$ term by term
\begin{equation}
\label{6.18}
I=
\!\!\!\!\!\!
\raisebox{-1.4cm}[1.5cm][1.2cm]{\epsfysize 2.9cm\epsfbox{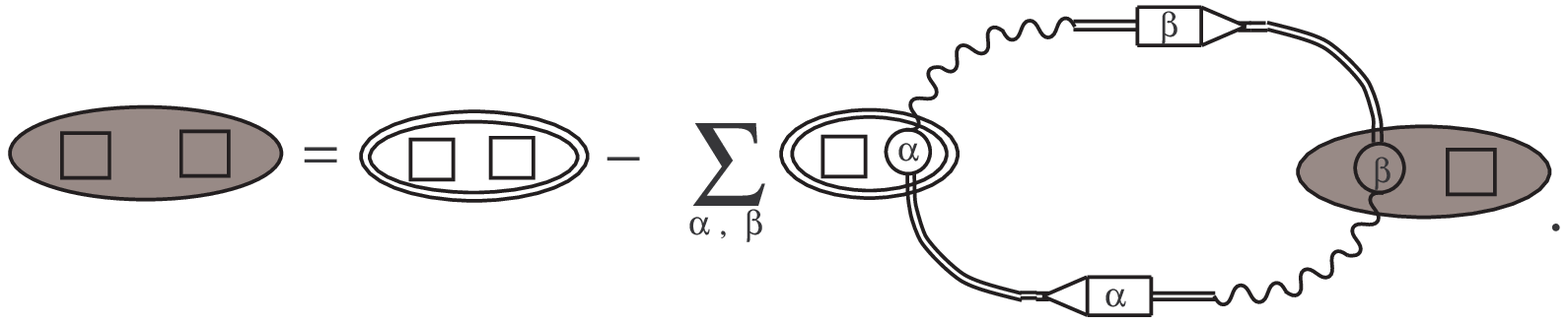}}
\hspace{-11mm}\null
\end{equation}
After simple transformation we can obtain the next relation
\begin{equation}
\label{6.19}
\langle nn \rangle_{MF}
-\langle n \rangle^2-
\frac{1}{2}\left(
\frac{\langle P^+\rangle}{\cosh^2\frac{\beta\varepsilon}{2}}+
\frac{\langle P^-\rangle}{\cosh^2\frac{\beta\tilde{\varepsilon}}{2}}
\right)=
\end{equation}
$$
=\frac{(
\langle n S^z \rangle_{MF}-\langle n \rangle\langle S^z \rangle)^2}
{\langle P^+ \rangle\langle P^-\rangle}\, .
$$
This relation makes possible to write immediately the simple analytic
expression for series $(\ref{6.18})$
\begin{equation}
\label{6.20} I=\Bigg\{
[2(n(\varepsilon)-n(\tilde{\varepsilon}))]^2
\langle S^zS^z\rangle_{\vec{q}}+\frac{1}{2}\left(
\frac{\langle P^+\rangle}{\cosh^2\frac{\beta\varepsilon}{2}}+
\frac{\langle P^-\rangle}{\cosh^2\frac{\beta\tilde{\varepsilon}}{2}}
\right)\Bigg\}\delta(\omega)\, .
\end{equation}
Analytical expressions for $I\! I$--term can be obtained starting from
formulae $(\ref{6.11})$ -- $(\ref{6.15})$
\begin{equation}
\label{6.21} I\! I
=\big\{2[n(\varepsilon)-n(\tilde{\varepsilon})]\langle
S^zS^z\rangle_{\vec{q}}\times [\oplus]_{\vec{q}}\big\}\delta(\omega)\, .
\end{equation}
Using the expression $(\ref{6.16})$ we can unite $(\ref{6.21})$ and
$(\ref{6.20})$
\begin{equation}
\label{6.22} I+I\! I=
\Bigg\{2(n(\varepsilon)-n(\tilde{\varepsilon}))\langle S^z n\rangle_{\vec{q}}
+\frac{1}{2}\left(
\frac{\langle P^+\rangle}{\cosh^2\frac{\beta\varepsilon}{2}}+
\frac{\langle P^-\rangle}{\cosh^2\frac{\beta\tilde{\varepsilon}}{2}}
\right)\Bigg\}\delta(\omega)\, .
\end{equation}
The diagram series for the fourth term in $(\ref{6.17})$ has form
$$
{\epsfysize 5.5cm\epsfbox{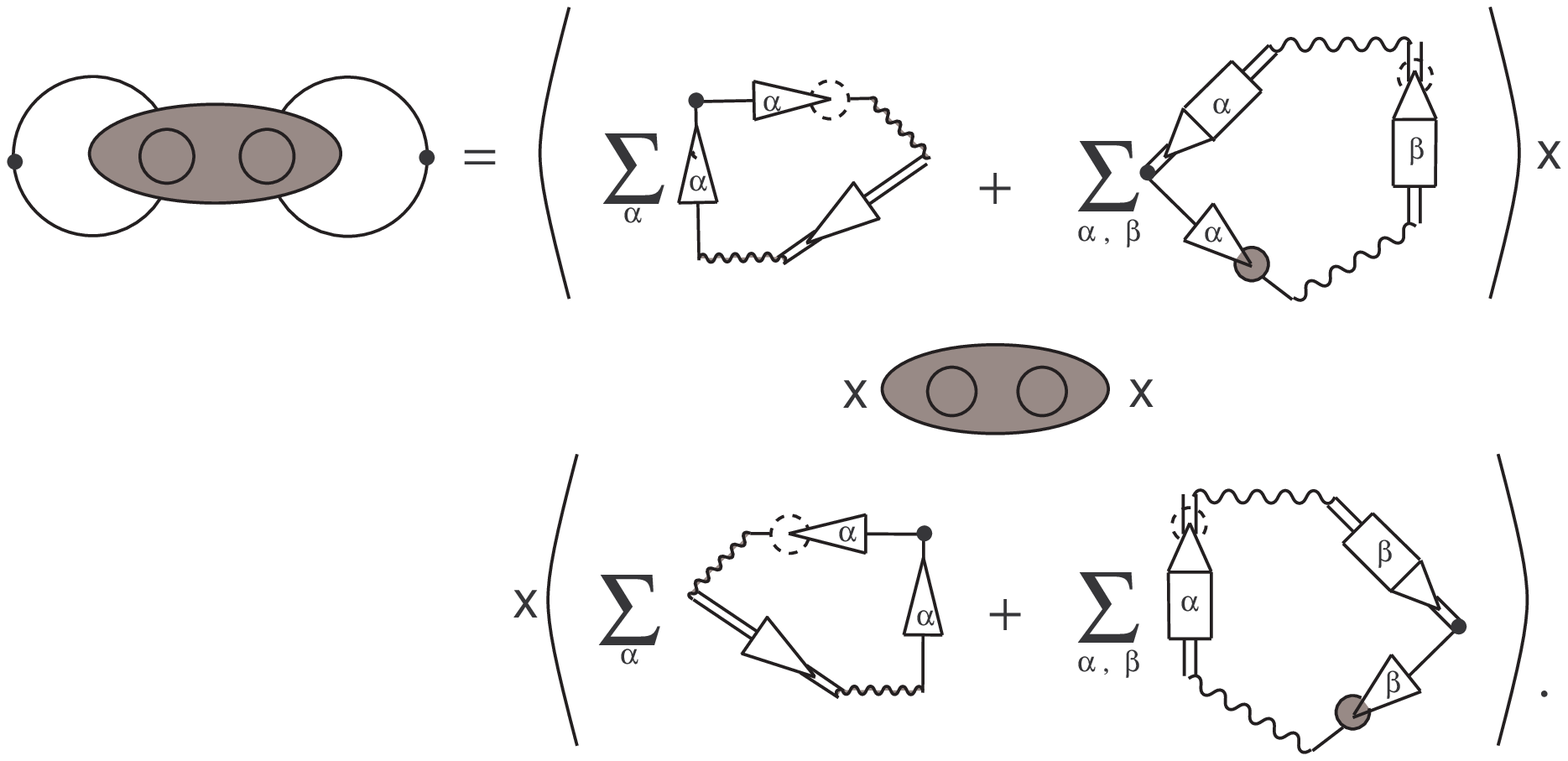}}
$$
And can be written as
\begin{equation}
\label{6.23}
I\! V=
\!\!\!\!\!
\raisebox{-.8cm}[1.2cm][.8cm]{\epsfysize 2.1cm\epsfbox{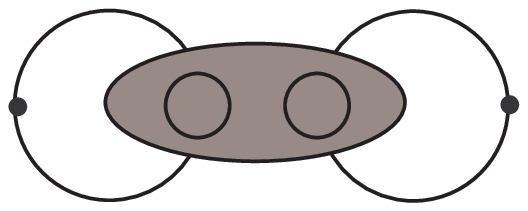}}
\hspace{-3em}
=[\oplus]_{\vec{q}}\times \langle S^z
S^z\rangle_{\vec{q}}\times [\oplus]_{\vec{q}}\cdot\delta(\omega)\, .
\end{equation}
Ones more using the formula $(\ref{6.16})$ we unite $I\! I\! I\!$--term and
$I\! V$--term
\begin{equation}
\label{6.24}
I\! I\! I\! +I\! V
=\langle n S^z\rangle_{\vec{q}}\times [\oplus]_{\vec{q}}
\cdot\delta(\omega)\, .
\end{equation}
Last term can be presented in the form
\begin{equation}
\raisebox{-1.4cm}[1.4cm][1.4cm]{\epsfysize 2.8cm\epsfbox{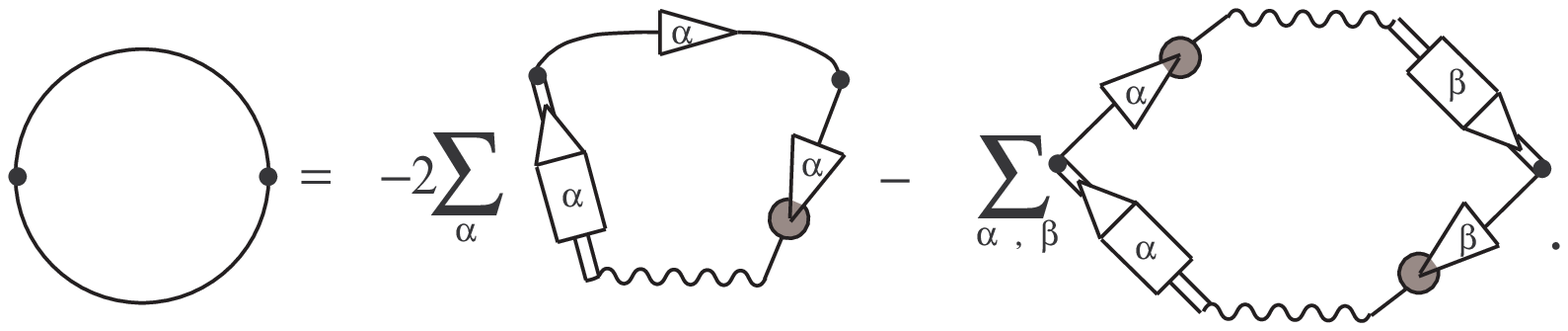}}
\end{equation}
Let us the take down final formula for electron correlator for the uniform
($\vec{q}=0$) and static ($\omega =0$) case
\begin{equation}
\hspace{-2em}
\langle nn \rangle=
2(n(\varepsilon)-n(\tilde{\varepsilon}))\langle S^z n\rangle_{\vec{q}=0}+
\frac{1}{2}\left(
\frac{\langle P^+\rangle}{\cosh^2\frac{\beta\varepsilon}{2}}+
\frac{\langle P^-\rangle}{\cosh^2\frac{\beta\tilde{\varepsilon}}{2}}
\right)+
\end{equation}
$$
\hspace{-1em}
\label{6.25}
+\frac{\beta}{2N}\sum_{\vec{k}}\frac
{t_{\vec{k}}(\varepsilon-\tilde{\varepsilon})}
{\varepsilon_I(t_{\vec{k}})-\varepsilon_{I\! I}(t_{\vec{k}})}
\left\{ \frac{1}{\cosh^2\frac{\beta\varepsilon_{I\! I}(t_{\vec{k}})}{2}}
-\frac{1}{\cosh^2\frac{\beta\varepsilon_{I}(t_{\vec{k}})}{2}}
\right\}\langle S^z n\rangle_{\vec{q}=0}+
$$
$$
+\frac{1}{2N}\sum_{\vec{k}}
\left\{ \frac{1}{\cosh^2\frac{\beta\varepsilon_{I\! I}(t_{\vec{k}})}{2}}
+\frac{1}{\cosh^2\frac{\beta\varepsilon_{I}(t_{\vec{k}})}{2}}
\right\}-\frac{1}{2}
\left\{ \frac{1}{\cosh^2\frac{\beta\varepsilon}{2}}
+\frac{1}{\cosh^2\frac{\beta\tilde{\varepsilon}}{2}}
\right\} .
$$
Same result we can obtain from the derivative
$ d\langle n \rangle/(d\beta\mu).$
Thus all our quantities: mean values of the pseudospin and particle number
operators, thermodynamical potential as well as correlation functions are
derived in the framework of one approximation which corresponds to the
mean field approximation.

In the second part of the paper we shell perform numerical calculations for
the analytical expressions obtained in the first part. We shall investigate
values of pseudospin and particle number operators with the change of the
asymmetry parameter $h$ ($T=const$) or with the change of temperature $T$
($h=const$) for the cases of the fixed chemical potential value
(regime $\mu =const$) and constant mean value particle number.

\end{document}